\documentclass[aps,prd,twocolumn,showpacs,amsmath,floatfix]{revtex4}
\usepackage[dvips]{graphicx}
\usepackage{bm}

\def\bea{\begin{eqnarray}}
\def\eea{\end{eqnarray}}
\def\be{\begin{equation}}
\def\ee{\end{equation}}
\def\rra{\right\rangle}
\def\lla{\left\langle}
\def\eps{\epsilon}
\def\sgm{\Sigma^-}
\def\la{\Lambda}
\def\piv{\bm{\pi}}
\def\kv{\bm{k}}
\def\zv{\bm{0}}

\begin{document}
\title{Hybrid stars with the color dielectric and the MIT bag models}
\author{C. Maieron, M. Baldo, G. F. Burgio, and H.-J. Schulze}
\affiliation{ 
INFN, Sezione di Catania, Via Santa Sofia 64, 95123 Catania, Italy}

\date{\today} 

\begin{abstract}
We study the hadron-quark phase transition in the interior of neutron
stars (NS). 
For the hadronic sector, we use a microscopic equation of state (EOS) 
involving nucleons and hyperons
derived within the Brueckner-Bethe-Goldstone many-body theory, 
with realistic two-body and three-body forces. 
For the description of quark matter, we employ both the MIT bag model 
with a density dependent bag constant, and the color dielectric model. 
We calculate the structure of NS interiors with the EOS comprising both 
phases, and we find that the NS maximum masses are never larger than 
1.7 solar masses, 
no matter the model chosen for describing the pure quark phase.
\end{abstract}

\maketitle

\section{Introduction}

The appearence of quark matter in the interior of massive neutron stars
(NS) is one of the main issues in the physics of these compact objects.
Calculations of NS structure, based on a microscopic nucleonic
equation of state (EOS), indicate that for the heaviest NS, close
to the maximum mass (about two solar masses), the central particle density
reaches values larger than $1/\rm fm^{3}$. 
In this density range the nucleon cores (dimension $\approx 0.5\;\rm fm$) 
start to touch each other, and it is hard to imagine that only 
nucleonic degrees of freedom can play a role. 
On the contrary, it can be expected that even before reaching 
these density values, the nucleons start to loose their identity, and quark 
degrees of freedom are excited at a macroscopic level. 

Unfortunately it is not straightforward to predict the relevance 
of quark degrees of freedom in the interior of NS for the different
physical observables, like cooling evolution, glitch characteristics,
neutrino emissivity, and so on. 
In fact, the other NS components can mask the effects coming directly 
from quark matter.
In some cases the properties of quark and nucleonic matter are not very
different, and a clear observational signal of the presence 
of the deconfined phase inside a NS is indeed hard to find. 

The value of the maximum mass of NS is probably one of the physical 
quantities that are most sensitive to the presence of quark matter in NS. 
If the quark matter EOS is quite soft, the quark component is expected
to appear in NS and to affect appreciably the maximum mass value.
In this case the maximum mass is expected to be slightly
larger than the observational limit
(1.44 solar masses of the so-called Taylor pulsar \cite{taylor}).   
The observation of a large NS mass 
(larger than 2 solar masses) 
would imply, on the contrary, that the EOS of NS matter is stiff enough 
to keep the maximum mass at this large values. 
Purely nucleonic EOS are able to accomodate masses comparable
with these large values \cite{gle,bbb,akma}.
Since the presence of non-nucleonic degrees of freedom, 
like hyperons and quarks, tends usually to soften considerably the EOS 
with respect to purely nucleonic matter, thus lowering the mass value, 
their appearence would be incompatible with observations. 
The large value of the mass could then be explained only if 
both hyperonic and quark matter EOS are stiffer than expected.
In particular, the quark EOS should be assumed to be stiff enough to render the
deconfined phase energetically disfavoured.

In this paper we will discuss this issue in detail. 
Unfortunately, while the microscopic theory of the nucleonic EOS has 
reached a high degree of sophistication, the quark matter EOS is poorly 
known at zero temperature and at the high baryonic density 
appropriate for NS.
One has, therefore, to rely on models of quark matter, which contain
a high degree of uncertainity. 
The best one can do is to compare the predictions of different models 
and to estimate the uncertainty of the results for the NS matter as well 
as for the NS structure and mass.
In this paper we will use a definite nucleonic EOS, which has been developed
on the basis of nuclear matter many-body theory, and two different models
for the quark EOS, and compare the results. 
Confrontation with previous calculations shall also be discussed.

The paper is organized as follows.
In section \ref{s:bhf} we review the determination of the baryonic
EOS comprising nucleons and hyperons in the Brueckner-Hartree-Fock
approach.
Section \ref{s:qm} concerns the quark matter EOS according to the 
MIT bag model and the color dielectric model (CDM).
In section \ref{s:res} we present the results regarding neutron star structure
combining the baryonic and quark matter EOS for beta-stable nuclear matter.
Section \ref{s:end} contains our conclusions.

\section{Brueckner theory}
\label{s:bhf}

\subsection{EOS of nuclear matter}

The Brueckner-Bethe-Goldstone (BBG) theory is based on a linked cluster 
expansion of the energy per nucleon of nuclear matter 
(see Ref.~\cite{book}, chapter 1 and references therein).  
The basic ingredient in this many-body approach is the Brueckner reaction 
matrix $G$, which is the solution of the Bethe-Goldstone equation 
\be
 G[\rho;\omega] = v + \sum_{k_a, k_b} v 
 { \left|k_a k_b\rra Q \lla k_a k_b\right|
 \over \omega - e(k_a) - e(k_b) } G[\rho;\omega] \:, 
\label{e:bg}
\ee
where $v$ is the bare nucleon-nucleon (NN) interaction, $\rho$ is the nucleon 
number density, and $\omega$ the starting energy.  
The single-particle energy $e(k)$ 
(assuming $\hbar$=1 here and throughout the paper),
\be
 e(k) = e(k;\rho) = {{k^2}\over {2m}} + U(k;\rho),
\label{e:en}
\ee
and the Pauli operator $Q$ determine the propagation of intermediate 
baryon pairs. 
The Brueckner-Hartree-Fock (BHF) approximation for the 
single-particle potential
$U(k;\rho)$ using the continuous choice is
\be
 U(k;\rho) = {\rm Re} \sum _{k' < k_F} 
 \big\langle k k'\big| G[\rho; e(k)+e(k')] \big|k k'\big\rangle_a \:,
\ee
where the subscript ``$a$'' indicates antisymmetrization of the 
matrix element.  
Due to the occurrence of $U(k)$ in Eq.~(\ref{e:en}), they constitute 
a coupled system that has to be solved in a self-consistent manner
for several Fermi momenta of the particles involved. 
In the BHF approximation the energy per nucleon is
\be
 {E \over A} = {3\over5}{k_F^2\over 2m} + {1\over 2\rho}  
 \sum_{k,k' < k_F}  
 \big\langle k k'\big|G[\rho; e(k)+e(k')]\big|k k'\big\rangle_a. 
\ee
In this scheme, the only input quantity we need is the bare NN interaction
$v$ in the Bethe-Goldstone equation (\ref{e:bg}). 
In this sense the BBG approach can be considered as a microscopic one. 
The nuclear EOS can be calculated with good accuracy in the Brueckner 
two hole-line approximation with the continuous choice for the 
single-particle potential, since the results in this scheme are 
quite close to the calculations which include also the three hole-line
contribution \cite{song}. 
In the calculations reported here, we have used the Argonne $v_{18}$
potential \cite{v18} as the two-nucleon interaction.

However, it is commonly known that  
nonrelativistic calculations, based on purely two-body interactions, fail 
to reproduce the correct saturation point of symmetric nuclear matter,
and three-body forces (TBF) among nucleons are needed to correct this
deficiency. 
In this work the so-called Urbana model will be used, 
which consists of an attractive term due to two-pion exchange
with excitation of an intermediate $\Delta$ resonance, and a repulsive 
phenomenological central term \cite{schi}. 
We introduced the same Urbana three-nucleon
model within the BHF approach 
(for more details see Ref.~\cite{bbb}).
In our approach the TBF is reduced to a density dependent two-body force by
averaging over the position of the third particle, assuming that the
probability of having two particles at a given distance is reduced 
according to the two-body correlation function determined self-consistently. 
The corresponding nuclear matter EOS fulfills several requirements, namely 
(i) it reproduces the correct nuclear matter saturation point,
(ii) the incompressibility is compatible
with the values extracted from phenomenology, 
(iii) the symmetry energy is compatible with nuclear phenomenology, 
(iv) the causality condition is always fulfilled.  

In order to study the structure of neutron stars, we have to calculate 
the composition and the EOS of cold, neutrino-free, catalyzed matter. 
We require that the neutron star contains charge neutral matter 
consisting of neutrons, protons, and leptons ($e^-$, $\mu^-$)
in beta equilibrium, and compute the EOS for 
charge neutral and beta-stable matter
in the following standard way \cite{bbb,shapiro,gle}:
The Brueckner calculation yields the energy density of 
lepton/baryon matter as a function of the different partial densities, 
\bea
 \eps(\rho_n,\rho_p,\rho_e,\rho_\mu) &=& 
 (\rho_n m_n +\rho_p m_p) 
 + (\rho_n+\rho_p) {E\over A}(\rho_n,\rho_p)
\nonumber\\ && 
 +\, \rho_\mu m_\mu + {1\over 2m_\mu}{(3\pi^2\rho_\mu)^{5/3} \over 5\pi^2}
\nonumber\\ && 
 +\, { (3\pi^2\rho_e)^{4/3} \over 4\pi^2} \:,
\label{e:epsnn}
\eea
where we have used ultrarelativistic and nonrelativistic approximations
for the energy densities of electrons and muons, respectively.
In practice, it is sufficient to compute only the binding energy of
symmetric nuclear matter and pure neutron matter,
since within the BHF approach it has been verified \cite{hypns,bom} 
that a parabolic approximation for the binding
energy of nuclear matter with arbitrary proton fraction 
$x=\rho_p/\rho$, $\rho=\rho_n+\rho_p$,
is well fulfilled,
\be
 {E\over A}(\rho,x) \approx 
 {E\over A}(\rho,x=0.5) + (1-2x)^2 E_{\rm sym}(\rho) \:,
\label{e:parab}
\ee
where
the symmetry energy $E_{\rm sym}$ can be expressed in
terms of the difference of the energy per particle between pure neutron 
($x=0$) and symmetric ($x=0.5$) matter:
\be
  E_{\rm sym}(\rho) = 
  - {1\over 4} {\partial(E/A) \over \partial x}(\rho,0)
  \approx {E\over A}(\rho,0) - {E\over A}(\rho,0.5) \:.
\label{e:sym}
\ee

Knowing the energy density Eq.~(\ref{e:epsnn}), 
the various chemical potentials (of the species $i=n,p,e,\mu$)
can be computed straightforwardly,
\be
 \mu_i = {\partial \eps \over \partial \rho_i} \:,
\ee
and the equations for beta-equilibrium,
\be
\mu_i = b_i \mu_n - q_i \mu_e \:,
\ee 
($b_i$ and $q_i$ denoting baryon number and charge of species $i$)
and charge neutrality,
\be 
 \sum_i \rho_i q_i = 0 \:,
\ee
allow one to determine the
equilibrium composition $\rho_i(\rho)$
at given baryon density $\rho$ 
and finally the EOS,
\be
 P(\rho) = \rho^2 {d\over d\rho} 
 {\eps(\rho_i(\rho))\over \rho}
 = \rho {d\eps \over d\rho} - \eps 
 = \rho \mu_n - \eps \:.
\ee

\subsection{Hyperons in nuclear matter}

While at moderate densities $\rho \approx \rho_0$ the matter inside 
a neutron star consists only of nucleons and leptons, 
at higher densities several other
species of particles may appear due to the fast rise of the baryon 
chemical potentials with density. 
Among these new particles are strange baryons, namely, 
the $\Lambda$, $\Sigma$, and $\Xi$ hyperons. 
Due to its negative charge, the $\Sigma^-$ hyperon is the 
first strange baryon expected to appear with increasing density in the 
reaction $n+n \rightarrow p+\Sigma^-$,
in spite of its substantially larger mass compared to the neutral 
$\Lambda$ hyperon 
($M_{\Sigma^-}=1197\;{\rm MeV}, M_\Lambda=1116\;{\rm MeV}$).
Other species might appear in stellar matter,
like $\Delta$ isobars along with pion and kaon condensates.
It is therefore mandatory to generalize the study of the nuclear EOS
with the inclusion of the possible hadrons, other than nucleons, which
can spontaneously appear in the inner part of a neutron star, 
just because their appearance is able to lower the ground state energy 
of the dense nuclear matter phase. 

As we have pointed out in the previous section, the nuclear EOS can be 
calculated with good accuracy in the Brueckner two hole-line 
approximation with the continuous choice for the single-particle
potential, since the results in this scheme are quite close to the full
convergent calculations which include also the three hole-line
contribution. 
It is then natural to include the hyperon degrees of freedom
within the same approximation to calculate the nuclear EOS needed
to describe the neutron star interior. 
To this purpose, one requires in principle 
nucleon-hyperon (NY) and hyperon-hyperon (YY) potentials. 
In our work we use the Nijmegen soft-core NY potential \cite{mae89}
that is well adapted to the available experimental NY scattering data.
Unfortunately, up to date no YY scattering data 
and therefore no reliable YY potentials are available.
We therefore neglect these interactions in our calculations,
which is supposedly justified, as long as the hyperonic partial 
densities remain limited.
Also, for the following calculations the $v_{18}$ NN potential together
with the phenomenological TBF introduced previously, are used.

With the NN and NY potentials, the various $G$ matrices are evaluated 
by solving numerically the Brueckner equation, which can be written in 
operatorial form as \cite{hypmat,hypns}
\be
  G_{ab}[W] = V_{ab} + \sum_c \sum_{p,p'} 
  V_{ac} \big|pp'\big\rangle 
  { Q_c \over W - E_c +i\eps} 
  \big\langle pp'\big| G_{cb}[W] \:, 
\label{e:g}
\ee
where the indices $a,b,c$ indicate pairs of baryons
and the Pauli operator $Q$ and energy $E$ 
determine the propagation of intermediate baryon pairs.
In a given nucleon-hyperon channel $c=(NY)$ one has, for example,
\be
  E_{(NY)} = m_N + m_Y + {k_N^2\over2m_N} + {k_Y^2\over 2m_Y} +
  U_N(k_N) + U_Y(k_Y) \:.
\label{e:e}
\ee
The hyperon single-particle potentials within the continuous choice
are given by
\bea
 U_Y(k) &=& {\rm Re} \sum_{N=n,p}\sum_{k'<k_F^{(N)}} 
\nonumber\\ && \times
 \Big\langle k k' \Big| G_{(NY)(NY)}\big[E_{(NY)}(k,k')\big] 
 \Big| k k' \Big\rangle \qquad
\label{e:uy}
\eea
and similar expressions of the form
\be
 U_N(k) = \sum_{N'=n,p} U_N^{(N')}(k) + 
 \sum_{Y=\Sigma^-,\Lambda} U_N^{(Y)}(k) 
\label{e:un}
\ee
apply to the nucleon single-particle potentials.
The nucleons feel therefore direct effects of the other nucleons as well as 
of the hyperons in the environment, whereas for the hyperons there are only 
nucleonic contributions, because of the missing hyperon-hyperon potentials.
The equations (\ref{e:g})--(\ref{e:un}) define the BHF scheme with the 
continuous choice of the single-particle energies.  
In contrast to the standard purely nucleonic calculation
there is now an additional coupled channel structure,
which renders a self-consistent calculation quite time-consuming.
 
Once the different single-particle potentials are known,
the total nonrelativistic baryonic energy density, $\eps$,  
can be evaluated:
\begin{eqnarray}
 \eps &=&\sum_{i=n,p,\Sigma^-,\Lambda} \int_0^{k_F^{(i)}}
 {dk\,k^2\over\pi^2} 
 \left[ m_i + {k^2\over{2m_i}} + {1\over2}U_i(k) \right] \qquad
\\
 &=& \eps_{NN} +
 \sum_{Y=\sgm,\la} 
 \int_0^{k_F^{(Y)}} {dk\,k^2\over\pi^2}
\nonumber\\ && \times
 \left[ m_Y + {k^2\over 2m_Y} + 
 U_Y^{(n)}(k) + U_Y^{(p)}(k) \right] \:, \qquad
\label{e:epsny}
\eea
where $\eps_{NN}$ is the nucleonic part of the energy density, 
Eq.~(\ref{e:epsnn}).
Using for example an effective mass approximation for the hyperon 
single-particle potentials, one could write the last term due to the 
nucleon-hyperon interaction as
\be
 \eps_{NY} = \sum_{Y=\sgm,\la} \left(
 \rho_Y \Big[ m_Y + U_Y(0) \Big] +
 {1\over 2m_Y^*}{(3\pi^2\rho_Y)^{5/3} \over 5\pi^2} \right) \:,
\ee
which should be added to Eq.~(\ref{e:epsnn}).

The knowledge of the energy density  
allows one then to compute EOS and neutron star structure as described before, 
now making allowance for the species $i=n,p,\sgm,\la,e^-,\mu^-$.
The main physical features of the nuclear EOS which determine the
resulting compositions are essentially the symmetry energy of the nucleon 
part of the EOS and the hyperon single-particle potentials inside 
nuclear matter. 
Since at low enough density the nucleon matter is quite 
asymmetric, the small percentage of protons feel a deep single-particle
potential, and therefore it is energetically convenient to create 
a $\Sigma^{-}$ hyperon, since then a neutron can be converted into a proton.
The depth of the proton potential is mainly determined by the
nuclear matter symmetry energy. 
Furthermore, the potentials felt by the
hyperons can shift substantially the threshold density at which each
hyperon sets in. 

We have found rather low hyperon onset densities of about 2 to 3 times 
normal nuclear matter density
for the appearance of the $\sgm$ and $\la$ hyperons \cite{hypns}.
(Other hyperons do not appear in the matter).
Moreover, an almost equal percentage of nucleons and hyperons are 
present in the stellar core at high densities. 
The inclusion of hyperons produces an EOS which turns out to be
much softer than the purely nucleonic case.
The consequences for the structure of the neutron stars are dramatic.
In fact the presence of hyperons leads 
to a maximum mass for neutron stars of less than 1.3 solar masses \cite{hypns},
which is below the observational limit.

This surprising result is due to the strong softening of the baryonic
EOS when including hyperons as additional degrees of freedom.
We do not expect substantial changes when introducing refinements
of the theoretical framework, such as 
hyperon-hyperon potentials \cite{barc}, relativistic corrections, etc. 
Three-body forces involving hyperons could produce a substantial stiffening 
of the baryonic EOS. 
Unfortunately they are essentially unkown, but can
be expected to be weaker than in the non-strange sector. 	 
Another possibility that is able to produce larger maximum masses,
is the appearence of a transition to another phase 
of dense (quark) matter inside the star. 
This will be discussed in the following.

\section{Quark Phase}
\label{s:qm}

The results obtained with a purely baryonic EOS call for an estimate of
the effects due to the hypothetical presence of quark matter in the interior
of the neutron star.
Unfortunately, the current theoretical description of quark matter 
is burdened with large uncertainties, seriously limiting the 
predictive power of any theoretical approach at high baryonic density. 
For the time being we can therefore only resort
to phenomenological models for the quark matter EOS 
and try to constrain them as well as possible 
by the few experimental information on high density baryonic matter.

One of these constraints is the phenomenological
observation that in heavy ion collisions at intermediate energies
($10\;{\rm MeV}/A \lesssim E/A \lesssim 200\;{\rm MeV}/A$) 
no evidence for a transition to a quark-gluon plasma has been found. 
Indeed, all microscopic simulations, like BUU or QMD, 
that are able to reproduce a great variety of experimental
data, do not need the introduction of such a transition \cite{daniele}. 
In these simulations the calculated nucleon density can reach values 
which are at least 2 to 3 times larger than the saturation density $\rho_0$.
One can, therefore, conclude that symmetric or nearly symmetric nuclear
matter at a few MeV of temperature does not exhibit any phase transition
to deconfined matter up to this baryon density. 
It has to be noticed that the phase transition in symmetric matter 
can occur at a substantially different baryon density than in 
neutron star matter,
where nuclear matter is closer to neutron matter than to symmetric
nuclear matter. 

This constraint coming from heavy-ion physics appears
as an independent one, that should be fulfilled by any theory or model
of deconfinement. 
Indeed, quark matter models
can have, in some cases, serious difficulties to fulfill the constraint 
(the transition occuring at too low density), 
even if they produce ``reasonable" results for neutron
stars, where the transition does occur.
We will in the following take this constraint in due consideration, 
and use an extended MIT bag model \cite{chodos}
(including the possibility of a density dependent bag ``constant'') 
and the color dielectric model \cite{Pirner84}, 
both compatible with this condition.

Furthermore, some theoretical interpretation of the heavy ion experiments 
performed at the CERN SPS \cite{cern}
points to a possible phase transition at a critical density
$\rho_c \approx 6\rho_0 \approx 1/{\rm fm}^3$.
This condition has to be taken with caution, since in ultrarelativistic
heavy ion collisions, quark matter is probed essentially at zero baryonic
density and high temperature, at variance with the case of NS matter,
where high baryonic density and essentially zero temperature are present.

\subsection{The MIT bag model}

We first review briefly the description of the bulk properties of 
uniform quark matter,
deconfined from the $\beta$-stable hadronic matter mentioned in the
previous section, by using the MIT bag model \cite{chodos}.
The thermodynamic potential of $f=u, d, s$ quarks 
can be expressed as a sum of the kinetic term
and the one-gluon-exchange term \cite{quark,fahri}
proportional to the QCD fine structure constant $\alpha_s$, 
\bea
 \Omega_f(\mu_f) &=& -{3m_f^4 \over 8\pi^2} \bigg[ 
 {y_f x_f \over 3} \left(2x_f^2-3\right) + \ln(x_f+y_f) \bigg] 
\nonumber \\
 &+& \alpha_s{3m_f^4\over 2\pi^3} \bigg\{
 \Big[ y_f x_f - \ln(x_f+y_f) \Big]^2 
 - {2\over3} x_f^4 + \ln(y_f) 
\nonumber \\
 &+& 2\ln\Big( {\sigma_{\rm ren} \over m_f y_f} \Big) \Big[ 
 y_f x_f - \ln(x_f + y_f) \Big] \bigg\} \:,
\eea
where $m_f$ and $\mu_f$ are the $f$ current quark mass and 
chemical potential, respectively, 
$y_f = \mu_f/m_f$, $x_f = \sqrt{y_f^2-1}$,
and
$\sigma_{\rm ren}$ = 313~MeV
is the renormalization point. 
In this work we will consider massless $u$ and $d$ quarks
(together with $m_s=150\;\rm MeV$),
in which case the above expression reduces to 
\be
 \Omega_q = -{\mu_q^4 \over 4\pi^2} 
 \Big(1 - {2\alpha_s \over \pi} \Big) \:, 
 \qquad (q=u,d) \:.
\ee
The number density $\rho_f$ of $f$ quarks is related to $\Omega_f$ via
\be
 \rho_f = - {\partial\Omega_f \over \partial\mu_f} \:,
\ee
and the total energy density for the quark system is written as
\be
 \eps_{\rm MIT}(\rho_u,\rho_d,\rho_s) = 
 \sum_f \big( \Omega_f + \mu_f \rho_f \big) + B \:,  
\label{e:eosqm}
\ee
where $B$ is the energy density difference between 
the perturbative vacuum and the true vacuum, i.e., the bag ``constant.''
In the original MIT bag model 
$B\approx 55\;\rm MeV\,fm^{-3}$ is used,
while values
$B\approx 210\;\rm MeV\,fm^{-3}$ 
are estimated from lattice calculations \cite{satz1}. 
In this sense $B$ can be considered as a free parameter.

The composition of $\beta$-stable quark matter is determined by 
imposing the condition of equilibrium under weak interactions for the
following processes
\begin{subequations}
\bea
 u + e^- & \rightarrow & d + \nu_e \:, \\
 u + e^- & \rightarrow & s + \nu_e \:, \\
 d & \rightarrow & u + e^- + \overline \nu_e \:, \\
 s & \rightarrow & u + e^- + \overline \nu_e \:, \\
 s + u & \rightarrow & d + u \:. 
\eea
\end{subequations}
In neutrino-free matter 
($\mu_{\nu_e} = \mu_{\overline \nu_e} = 0$),
the above equations imply for the chemical potentials
\be
 \mu_d = \mu_s = \mu_u + \mu_e \:.  
\label{eq:chem}
\ee
As in baryonic matter, the relations for chemical equilibrium must be 
supplemented with the charge neutrality condition and the total baryon 
number conservation,
\bea
 0 &=& {1\over 3}(2\rho_u - \rho_d - \rho_s) - \rho_e \:,
\\
 \rho &=& {1\over3} (\rho_u + \rho_d + \rho_s) \:, 
\label{eq:baryon}
\eea
in order to determine 
the composition $\rho_f(\rho)$ and the pressure of the quark phase,
\be
 P_Q(\rho) = \rho {d\eps_Q \over \ d\rho} - \eps_Q \:.
\ee

It has been found \cite{alford,qm} 
that within the MIT bag model
(without color superconductivity) 
with a density independent bag constant $B$, 
the maximum mass of a NS cannot exceed a value of about 1.6 solar masses. 
Indeed, the maximum mass increases as the value of $B$ decreases, 
but too small values of $B$ are incompatible 
with a transition density $\rho > (2,\ldots,3)\rho_0$ 
in symmetric nuclear matter,
as demanded by heavy-ion collision phenomenology 
according to the preceeding discussion. 
In order to avoid this serious drawback of the model,
one can introduce a density-dependent bag ``constant" $B(\rho)$, 
and this approach was followed in Refs.~\cite{qm}. 
This allows one to lower the value of $B$ at large density, 
providing a stiffer quark matter EOS and increasing
the value of the maximum mass, 
while at the same time still fulfilling the condition
of no phase transition below $\rho \approx 3 \rho_0$.
The comparison of the predictions based on the MIT bag model and the
CDM can be considered meaningful only if this constraint is
maintained also in the CDM.

In the following we present results based on the MIT model
using a constant value of the bag constant, 
$B=90\;\rm MeV/fm^3$,
and a gaussian parametrization for the density dependence,
\be
 {B(\rho)} = B_\infty + (B_0 - B_\infty)  
 \exp\left[-\beta\Big({\rho \over \rho_0}\Big)^2 \right]
\label{eq:param} 
\ee
with $B_\infty = 50\;\rm MeV\,fm^{-3}$, $B_0 = 400\;\rm MeV\,fm^{-3}$,
and $\beta=0.17$.

\subsection{The color dielectric model}

Let us now consider the color dielectric model, which was originally
introduced \cite{Pirner84} as a confinement model of the nucleon 
(see the general reviews \cite{Birse90,Pirner92,Banerjee93}).
In the CDM the nucleon is described as a soliton
in which quarks are dynamically confined via the interaction with a 
scalar-isoscalar chiral singlet field, indicated in the following as $\chi$, 
whose quanta correspond to glueballs or hybrid mesons.
Several closely related versions of the CDM
exist in the literature. 
They have been widely employed in the baryon sector,
to calculate the static 
\cite{Dodd:pw,McGovern91,Leech:bt,Neuber93} 
and dynamic properties 
\cite{Barone:uc,Barone:1996un,Barone:iq,Drago:1997uj,Fiolhais:uh,Alberto01} 
of the nucleon and to describe strange baryons 
\cite{McGovern91,Aoki,Nishikawa,Bae}.
The CDM has also been applied 
in the quark sector, to calculate the EOS of quark matter 
\cite{McGovern90,Broniowski90,Ghosh:1991wi,BaroneDrago,DragoNPA}
and to study the stability of strange quark matter
\cite{Alberico01}.
Applications of the CDM EOS for quark matter to the study of compact stars
have been considered by Ghosh et al.~\cite{Ghosh:1994hr} and by
Drago and collaborators, who studied the structure of hybrid stars
\cite{DragoPLB96,DragoPLB01} and the problem of supernova explosions
\cite{Drago97}, and, more recently by Malheiro et al.
\cite{Malheiro01,Malheiro03}
to study pure quark stars.

In this work we use the chiral version of the CDM 
\cite{Broniowski90,DragoNPA},
extended to include strange quarks 
\cite{McGovern90,DragoPLB96}, which, as shown in Ref.~\cite{DragoNPA},
describes reasonably well the nucleon, while giving,
with the same set of parameter values, a meaningful equation of state
for symmetric quark matter.
The Lagrangian of the model reads
\bea
 {\mathcal{L}} &=& 
 \sum_{f=u,d,s} i \overline{\psi}_f \gamma^\mu \partial_\mu \psi_f 
\nonumber\\ &&  
 + \sum_{f=u,d} {g_f \over f_\pi \chi} \overline{\psi}_f 
 \left( \sigma + i \bm{\tau}\cdot\piv\gamma_5 \right)\psi_f
 - {g_s \over \chi} \overline{\psi}_s \psi_s
\nonumber\\ &&  
 +\, {1\over2} \left(\partial_\mu \sigma \right)^2
 + {1\over2} \left(\partial_\mu \piv \right)^2
 - U(\sigma,\piv)
\nonumber\\ &&  
 +\, {1\over2} \left(\partial_\mu \chi \right)^2 - V(\chi) \:. 
\label{eq:CDMlag}
\eea
Here the potential $V(\chi)$ has the quadratic form
\be
 V(\chi) = {1\over2} M^2 \chi^2 \:, 
\label{eq:vchi}
\ee
while 
\be
 U(\sigma,\piv) = {m_\sigma^2 \over 8 f_\pi^2}
 \left(\sigma^2 + \piv^2 - f_\pi^2\right)^2
\label{eq:umex}
\ee
is the usual ``Mexican hat'' potential.
The characteristic feature of the CDM is the coupling of the quarks
to an inverse power of the field $\chi$, through which the quarks
acquire density dependent effective masses 
$m_{u,d} = -{g_{u,d} \sigma}/{f_\pi \chi}$ and 
$m_s = {g_s}/{\chi}$.
Since $\chi$ vanishes in the vacuum, the quark masses diverge
as the density $\rho$ goes to zero, thus providing confinement.

In Eq.~(\ref{eq:CDMlag}) the couplings are given by
$g_{u,d} = g(f_\pi \pm \xi_3)$ and $g_s= g(2 f_K -f_\pi)$,
where $f_\pi=93$ MeV and $f_K=113$ MeV
are the pion and the kaon decay constants, respectively, and where 
$\xi_3=f_{K^\pm}-f_{K^0}= -0.75$ MeV. 
In Eq.~(\ref{eq:umex}) we take $m_\sigma=1.2$ GeV.

The parameters of the model are thus $g$, 
determining the couplings $g_f$, 
and the mass $M$ of the $\chi$ field. 
At the mean field level, and with the form Eq.~(\ref{eq:vchi})
of the potential, the only free parameter is actually the product 
$G=\sqrt{g M}$.
In the following we will use the values $M=1.7$ GeV and $g=23$ MeV
(corresponding to $G=198$ MeV), 
which were obtained in Ref.~\cite{DragoPLB96}, 
by requiring that the model provides reasonable values
for the average delta-nucleon mass and for the nucleon isoscalar radius. 

To describe the possible quark phase in the interior of a neutron star
we consider a uniform system of plane wave $u$, $d$, and $s$ quarks 
interacting with the fields $\chi$ and $\sigma$.
In the mean field approximation the latter are assumed to be constant, 
while the pion field vanishes.
The energy density of the system is then given by
\bea
 \epsilon_{\rm CDM} &=& 
 \gamma \sum_{f=u,d,s} \sum_{k<k_F^{f}}
 \sqrt{\kv^2+m_f(\sigma,\chi)^2}
\nonumber\\
 && +\, V(\chi) + U(\sigma,\piv=\zv) \:,
\label{eq:eqm}
\eea
where $\gamma=6$ is the spin and color degeneracy factor and
$k_F^f$  are the Fermi momenta of the quarks of flavor $f$.
At fixed baryon density, two coupled equations
for the fields $\chi$ and $\sigma$ are obtained by minimizing 
the energy density $\eps_{\rm CDM}$:
\bea
 \!\!\!\!\!\!\!&& {dV(\chi) \over d\chi} = 
 - \sum_{f=u,d} \rho_S^f(k_F^f,m_f) {g_f\sigma \over f_\pi \chi^2}
 + \rho_S^s(k_F^s,m_s) {g_s \over \chi^2} \:, \qquad
\label{eq:chi}
\\
 \!\!\!\!\!\!\!&& {dU(\sigma,\zv) \over d\sigma} = 
 \sum_{f=u,d} \rho_S^f(k_F^f,m_f) {g_f \over f_\pi \chi} \:, 
\label{eq:sigma}
\eea
where 
\be
 \rho_S^f(k_F^f,m_f) = \gamma \sum_{k<k_F^f}
 {m_f \over \sqrt{\kv^2 + m_f^2}}
\label{eq:scal}
\ee
are the quark scalar densities. 
By imposing chemical equilibrium, 
supplemented by the conditions of charge neutrality 
and baryon number conservation, 
Eqs.~(\ref{eq:chem})-(\ref{eq:baryon}) along with 
Eqs.~(\ref{eq:chi})-(\ref{eq:scal})  
form a system of six coupled equations, which are solved self-consistently 
to get $\chi,\sigma,k_F^u,k_F^d,k_F^s$, and $k_F^e$. 
Once they are solved, we obtain the population of each quark flavor
$\rho_f = \gamma(k_F^f)^3\! / 6\pi^2$,
as well as the one of the electrons.

\section{Results and discussion}
\label{s:res}

\subsection{Phase transition in symmetric matter}

\begin{figure}[t] 
\includegraphics[width=9cm]{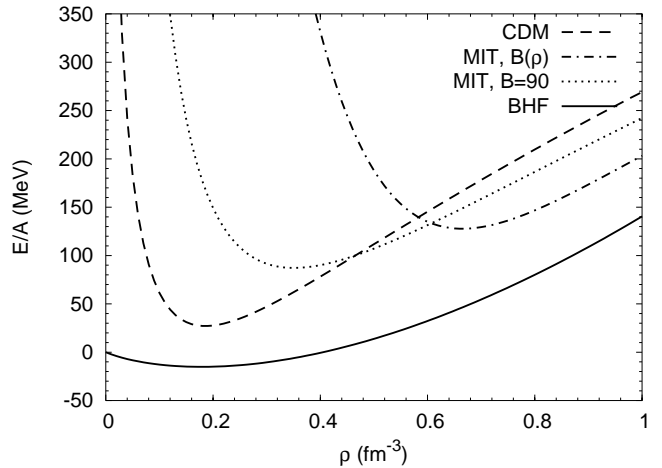}
\caption{
Energy per baryon for symmetric matter calculated
for the purely nucleonic case (solid line), 
with the MIT bag models (dotted and dot-dashed lines), 
and with the CDM (dashed line).}
\label{fig:EOSsym}
\end{figure} 

Before studying the quark phase inside neutron stars, let us first
discuss the EOS of symmetric matter.
We calculate the EOS for cold symmetric nuclear matter
in the BHF formalism with two-body and three-body forces, 
as described above in section~\ref{s:bhf}.
Then we calculate the EOS for $u$ and $d$ quark matter, using respectively 
Eq.~(\ref{e:eosqm}) for the MIT bag model and Eq.~(\ref{eq:eqm}) for the CDM. 
The results are displayed in Fig.~\ref{fig:EOSsym}.
The solid line represents the purely 
nucleonic case within the Brueckner many-body approach,
whereas the different broken lines denote the results for the 
various quark matter EOS.
We find that, in the range of baryon densities explored, 
the quark matter energy is always higher than that of 
symmetric nuclear matter, independently of the model 
adopted for describing the quark phase. 
Therefore nuclear matter is the favorite state. 
These results have been obtained choosing $\alpha_s=0$, but we have checked 
that they are not very sensitive to the value of $\alpha_s$. 

As far as the results obtained with the MIT bag model are concerned,
some clarifications are needed. 
In fact, apart from a bag constant $B=90\;\rm MeV\, fm^{-3}$,
we have also used a density dependent bag parameter $B(\rho)$, 
whose parametrization is the one of Eq.~(\ref{eq:param}).
This was adopted in previous works \cite{qm}, and we do not
repeat here the motivations and the technical details. 

The conclusion that indeed no transition to quark matter is present below 
$(3,\ldots,4)\rho_0$ can be considered well established. 
At higher baryonic density the precise trend of the EOS of symmetric 
hadronic matter is surely more uncertain. 
A refinement of the EOS could lead to a transition 
to a deconfined phase at a higher baryonic density, compatible with
the value extracted from the CERN SPS experimental data
($\rho_c \approx 6\rho_0 \approx 1/{\rm fm}^3$). 
In particular, different sets of three-body forces and the inclusion of 
three hole-line diagrams can produce indeed a stiffening of the hadronic 
EOS and the transition to quark matter can occur at this high density. 
A critical discussion on this point will be published elsewhere.
However, these refinements do not affect the results in NS matter, 
since there, due to the appearence of hyperons and quark matter, 
the purely hadronic phase never reaches a density higher 
than $(3,\ldots,4)\rho_0$, 
and the uncertainity in the high density EOS of nuclear matter 
plays no role. 

In order to facilitate the comparison between the different quark models 
adopted in this work, it is useful to introduce in the CDM 
an effective bag parameter, which describes the difference between the quark 
matter energy density and the energy density of a system of free quarks of
fixed mass $m'_f$ having the same density and composition.
We thus define
\be
 B_{\rm eff}(\rho) \equiv \epsilon_{\rm CDM}
 - \gamma \sum_{f=u,d,s} \sum_{k<k_F^f} \sqrt{\kv^2+{m_f'}^2} \:,
\label{eq:beff}
\ee 
where we take $m'_u=m_d'=0$, $m_s'=150\;\rm MeV$ in order to compare
with the bag parameter used in the MIT model. 
As an alternative \cite{Buballa04}, one can also define a bag parameter 
as the ``non-quark'' contribution to the energy density,
\bea
 B_{\rm eff}'(\rho) &\equiv& \epsilon_{\rm CDM}
 - \gamma \sum_{f=u,d,s} \sum_{k<k_F^f} 
 \sqrt{\kv^2+{m_f(\sigma,\chi)}^2} \qquad
\\
 &=& V(\chi) + U(\sigma,\piv=\zv) \:.
\label{eq:beff1} 
\eea 
The effective bag parameters $B_{\rm eff}$ and $B'_{\rm eff}$ are plotted
versus the baryon density $\rho$ in Fig.~\ref{fig:beff_ab}
for symmetric matter. 
We observe that, although at low baryon density the 
bag parameters used with the MIT model are larger than the
one calculated with the CDM, asymptotically they all reach values in the range 
$(50,\ldots,120)\;\rm MeV\;fm^{-3}$. 
However, in the former case the effective bag constant is a monotically 
decreasing function of the density, at variance with the CDM
and also with the Nambu-Jona-Lasinio model, widely studied in 
Refs.~\cite{njl,Buballa04}.    

\begin{figure} 
\includegraphics[width=9cm]{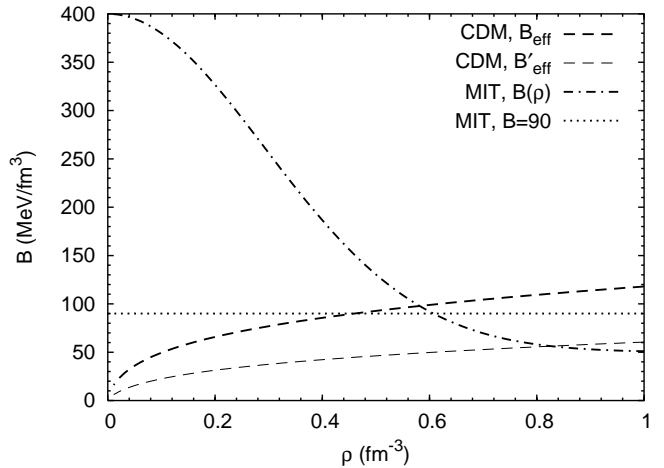}
\caption{
The bag parameters $B$ are shown as function of the
baryon density for symmetric matter. 
The dot-dashed line shows the parametrization 
Eq.~(\ref{eq:param}) used in the MIT model,
whereas the dashed lines represent the effective values of the CDM, 
calculated using Eqs.~(\ref{eq:beff}) or Eq.~(\ref{eq:beff1}), respectively.}
\label{fig:beff_ab}
\end{figure} 

\subsection{Phase transition in asymmetric $\beta$-stable matter}

\begin{figure*} 
\includegraphics[width=17cm]{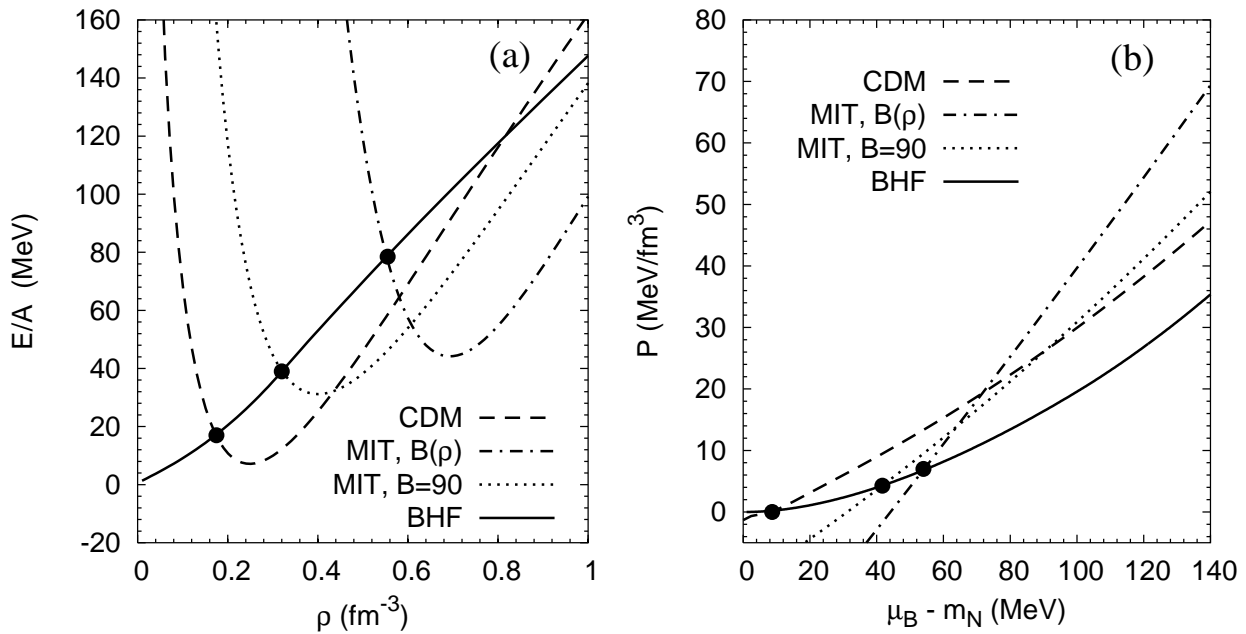}
\caption{
Panel (a) displays the energy per particle vs.~baryon density 
for beta-stable matter in the BHF approach (solid line) 
and for $u$,$d$,$s$ quark matter obtained within the 
MIT model (dotted and dot-dashed lines) and the CDM (dashed line).
Panel (b) shows the pressure as function of the baryonic chemical
potential for all cases.}
\label{fig:pchem_uds}
\end{figure*} 

We now consider the hadron-quark phase transition in neutron stars, and
calculate the EOS of a conventional neutron star as
composed of a chemically equilibrated and charge neutral mixture of
nucleons, hyperons, and leptons. 
The result is shown by the solid line in Fig.~\ref{fig:pchem_uds}(a).
The dashed line represents the EOS of beta-stable and
charge neutral ($u$,$d$,$s$) quark matter obtained within the CDM, 
and the dotted and dot-dashed lines the results due to the MIT model with 
constant and density dependent $B$, respectively.
Markers indicate the crossing points between the hadron and the quark phases. 
They lie inside the mixed phase region, whose range, at this stage, 
cannot be derived from the behavior of the energy alone. 
In spite of that, some qualitative considerations can be done. 
In particular, we notice that the phase transition from hadronic to quark matter
occurs at very low baryonic density when the CDM is used to describe 
the quark phase, whereas higher values of the transition density 
are predicted with the MIT bag model. 
In fact, the density dependent bag parameter was introduced in order to 
shift this transition to high density and to explore the implications for the 
NS observables.

A more realistic model for the phase transition between baryonic and quark phase 
inside the star is the Glendenning construction \cite{gle,glen},
which determines the range of baryon density where both phases coexist,
yielding an EOS containing
a pure hadron phase, a mixed phase, and a pure quark matter region.
In our previous papers \cite{qm}, we have used the Glendenning construction,
demonstrating that in particular the influence on 
the maximum mass value is rather small. 
Apart from that, the realization of the mixed phase depends on the nuclear
surface tension, which is currently an unknown parameter.

Therefore, in the present work we adopt the simpler Maxwell construction.
For that, we construct the phase transition from 
Fig.~\ref{fig:pchem_uds}(b), showing the pressure
as a function of the baryonic chemical potential $\mu_B$.
The coexistence region is determined by the intersection points
between the hadronic (solid line) and the different quark phases. 
We notice that the phase transition occurs at very low values of the baryon
density, $\rho \approx 0.05\;\rm fm^{-3}$, 
when the CDM is used.
The transition density that we obtain is somewhat lower than the value
obtained in Ref.~\cite{DragoPLB96}, where, using the Walecka model
to describe the hadronic EOS, the mixed phase was found to occur
in the interval 
$0.1\;\rm fm^{-3} \le \rho \le 0.31\;\rm fm^{-3}$. 
However, the general result
that the pure quark phase starts at a rather low density is in agreement
with Ref.~\cite{DragoPLB96}.

The phase transition constructed with the CDM is quite different
from the ones obtained using the MIT bag model.
In the latter case, the coexistence region is shifted to higher baryonic density,
as seen in Fig.~\ref{fig:pchem_uds}(b).
This implies a large difference in the structure of neutron stars.
In fact, whereas stars built with the CDM have at most a mixed phase
at low density and a pure quark core at higher density, the ones obtained
with the MIT bag model contain a hadron phase, followed by a mixed phase
and a pure quark interior. 
The scenario is again different 
within the Nambu-Jona-Lasinio model \cite{njl}, where at most a mixed phase
is present, but no pure quark phase.

\subsection{Neutron star structure}

We assume that a neutron star is a spherically symmetric distribution of 
mass in hydrostatic equilibrium. 
The equilibrium configurations are obtained
by solving the Tolman-Oppenheimer-Volkoff (TOV) equations \cite{shapiro} for 
the pressure $P$ and the enclosed mass $m$,
\begin{eqnarray}
  {dP(r)\over dr} &=& -{ G m(r) \epsilon(r) \over r^2 } \,
\nonumber\\ && \times
  {  \left[ 1 + {P(r) / \epsilon(r)} \right] 
  \left[ 1 + {4\pi r^3 P(r) / m(r)} \right] 
  \over
  1 - {2G m(r)/ r} } \:,\qquad
\\
  {dm(r) \over dr} &=& 4 \pi r^2 \epsilon(r) \:,
\end{eqnarray}
being $G$ the gravitational constant. 
Starting with a central mass density $\epsilon(r=0) \equiv \epsilon_c$,  
we integrate out until the pressure on the surface equals the one 
corresponding to the density of iron.
This gives the stellar radius $R$ and the gravitational mass is then 
\be
 M_G \equiv m(R) = 4\pi \int_0^R dr\; r^2 \epsilon(r) \:. 
\ee
We have used as input the equations of state discussed above for the
CDM and the MIT bag model for the beta-stable quark phase, and the BHF for 
the hadronic matter. 
For the description of the NS crust, we have joined 
the hadronic EOS with the ones by 
Negele and Vautherin \cite{nv} in the medium-density regime, and the ones   
by Feynman-Metropolis-Teller \cite{fey} and Baym-Pethick-Sutherland 
\cite{baym} for the outer crust. 

\begin{figure*} 
\includegraphics[width=9cm, angle=270]{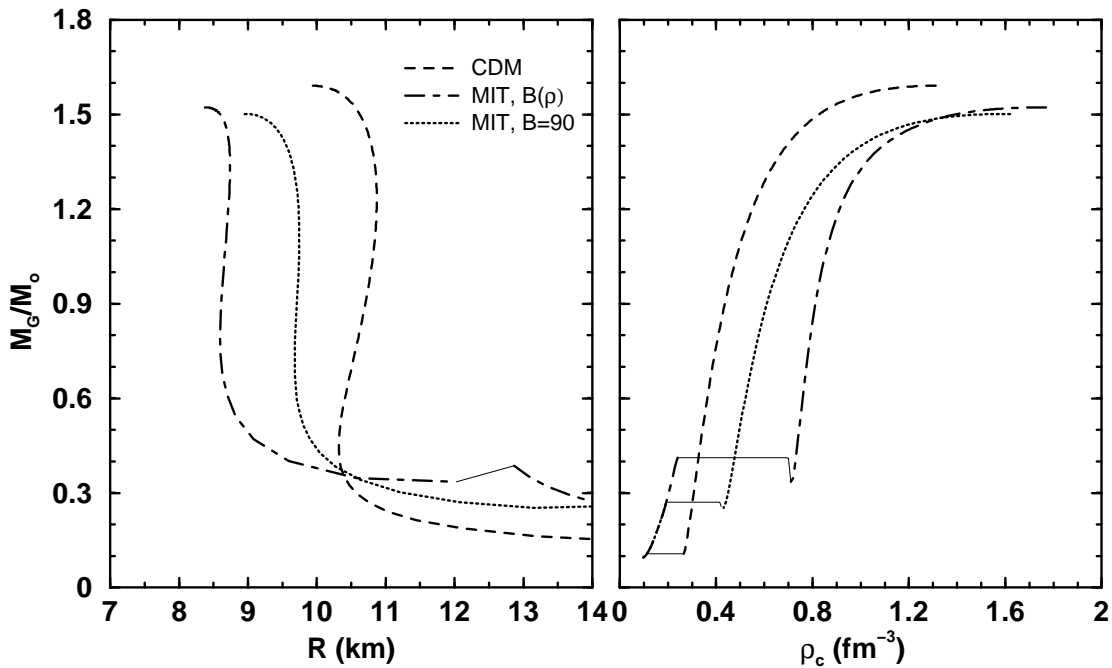}
\caption{
The mass (in units of solar mass $M_\odot=\rm 1.98\times10^{33}g$) 
is displayed as function of radius (left panel) 
and central density (right panel), using different EOS. 
See text for details.}
\label{fig:mr}
\end{figure*} 

The results are plotted in Fig.~\ref{fig:mr},
where we display the gravitational mass $M_G$ 
(in units of the solar mass $M_\odot$)
as a function of the radius $R$ (left panel) 
and central baryon density $\rho_c$ (right panel). 
The dashed lines represent the calculation
for beta-stable quark matter with the CDM, 
whereas the dotted and dot-dashed lines denote the
results obtained with the MIT bag model. 
Due to the use of the Maxwell construction, 
the curves are not continuous \cite{gle}:
For very small central densities (large radii, small masses) 
the stars are purely hadronic.
Then a sudden increase of the central density is required in order 
to start the quark phase in the center of the star, corresponding to the 
phase diagram Fig.~\ref{fig:pchem_uds}(b).
Due to this effect, also a discontinuity in the possible radii arises.
By performing the Glendenning construction, the curves would become continuous.

We observe that the values of the maximum mass depend only slighty 
on the EOS chosen for describing quark matter,
and lie between 1.5 and 1.6 solar masses.
A clear difference between the two models exists as far as the radius 
is concerned.
Hybrid stars built with the CDM are characterized by a larger radius
and a smaller central density, 
whereas hybrid stars constructed with the MIT bag model are more compact,
since they contain quark matter of higher density.

\begin{figure} 
\includegraphics[width=13cm,angle=270]{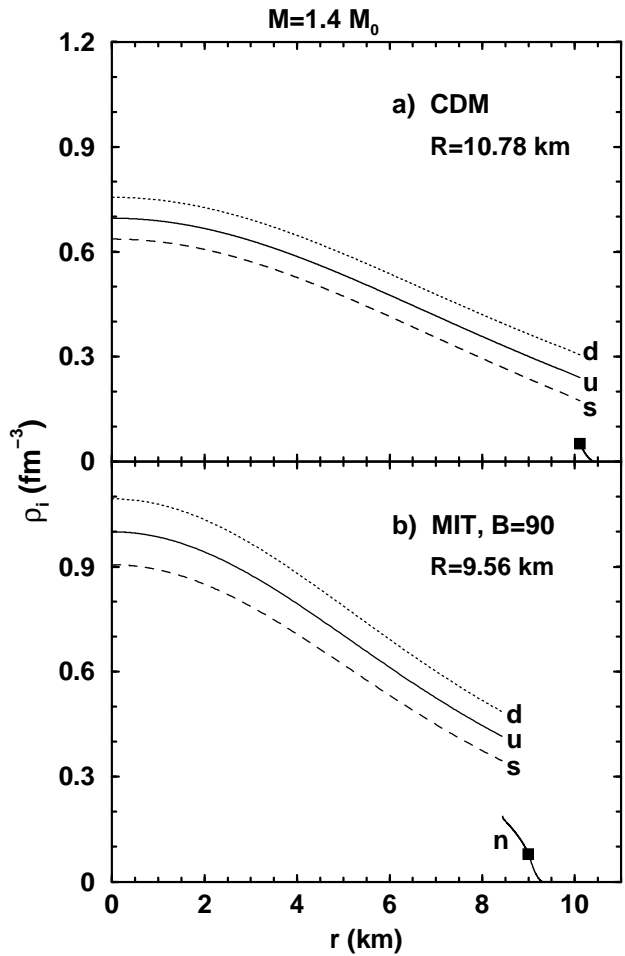}
\caption{
The particle populations inside a $M=1.4\,M_\odot$ neutron star 
with the CDM (upper panel) and the bag model (lower panel) quark matter EOS.
The markers indicate the beginning of the crust.
}
\label{fig:nr}
\end{figure} 

This is also illustrated in Fig.~\ref{fig:nr},
showing the different internal structure of stars with the 
MIT model and the CDM by comparing the populations of quarks and baryons 
inside a $M=1.4M_\odot$ neutron star.
At this value of the mass, within the MIT model only a thin hadronic layer 
is present,
mainly composed of neutrons, followed by a small portion of crust,
whereas in the case of the CDM one finds only crustal matter.
The central (quark) density is substantially larger with the MIT model,
in agreement with Fig.~\ref{fig:mr}.
In both models, the abrupt transition from quark to hadronic matter
is a consequence of the Maxwell construction.

\section{Conclusions}
\label{s:end}

In this article we determine the structure of neutron stars,
combining the most recent microscopic baryonic EOS in the BHF approach 
involving three-body forces and hyperons
with different effective models describing the quark matter phase.

Without allowing for the presence of quark matter, 
the maximum neutron star mass remains below 1.3 solar masses, 
due to the strong softening effect of the hyperons on the EOS,
compensating the repulsive character of nucleonic TBF at high density.
The presence of quark matter inside the star 
is required in order to reach larger maximum masses.

We introduced a density dependent bag parameter $B(\rho)$ in the MIT model
in order to explore the maximum NS mass that can be reached in this approach.
We compare with calculations using a fixed bag constant 
and using the color dielectric model.
Joining the corresponding EOS with the baryonic one, 
all three quark models yield
maximum masses in the range $(1.5,\ldots,1.6)\;M_\odot$,
while predicting slightly different radii.

Our results for the maximum masses are
in line with other recent calculations of neutron star properties employing
various phenomenological relativistic mean field nuclear EOS 
together with either effective mass bag model \cite{bag} 
or Nambu-Jona-Lasinio model \cite{njl} EOS for quark matter.

The value of the maximum mass of neutron stars obtained according to
our analysis appears rather robust with respect to the uncertainties
of the nuclear and the quark matter EOS.
Therefore, the experimental observation of a very heavy
($M \gtrsim 1.7 M_\odot$) neutron star, 
as claimed recently by some groups \cite{kaaret} 
($M \approx 2.2\;M_\odot$), 
if confirmed, would suggest that 
either serious problems are present for the current theoretical modelling
of the high-density phase of nuclear matter,
or that the assumptions about
the phase transition between hadron and quark phase
are substantially wrong. 
In both cases, one can expect a well defined hint on the
high density nuclear matter EOS.


\end{document}